# Intermolecular Interactions in Radial-Contour Mode Microring Resonators

Meysam T. Chorsi[1]

[1] Department of Mechanical Engineering, University of Connecticut, Storrs, CT 06269-3139, USA

**ABSTRACT**

This research is on the dynamics of electrostatically actuated radial-contour mode microring resonators. The governing equation of motion is derived by the minimization of the Hamiltonian and generalized to include the viscous damping effect. The Galerkin method is used to discretize the distributed-parameter model of the considered ring resonator. The influences of intermolecular forces such as van der Waals and Casimir on the dynamic behavior of the resonator are investigated. The natural frequencies and mode shapes of the ring are calculated for various values of ratio of radii ($\beta$). The effect of the design parameters including ring radius, electrostatic voltage and quality factor on the dynamic responses, is discussed. The results of present study can be used in the design of novel MEMS resonators, RF filters and channelizers.

**Keywords:** Microring; Resonator; Electrostatic force; Van der Waals force; Casimir force; Primary resonance.

## 1. INTRODUCTION

Microelectromechanical systems (MEMS) resonators have been widely used for signal filtering [1-8], biological sensing [9-16], mass sensing [17-19], timing references [20, 21], and other diverse applications.

MEMS resonators have many advantages such as high quality factor, small size, high sensitivity, and low cost batch fabrication. These advantages, together with the power advantages inherent in MEMS devices, make a compelling case for MEMS based resonators. MEMS resonators are typically based on three types of structures: microbeams [22-25], microdisks [26-31] and microrings [32-35]. As the size of the resonator is reduced, it has been observed that the attainable quality factors in beam resonators tend to decrease [36]. Microbeam resonators also suffer from large thermoelastic losses due to the bending motion [37]. On the other hand, radial-contour mode disk resonators can attain very high resonance frequency and still can retain relatively larger dimensions. The major bottleneck for utilization of disk resonators as RF components is still their extremely high motional resistance [38]. Pursuant to achieving high Q values at GHz frequencies with lower motional resistance, Li et al. [39] introduced the first radial-contour mode ring resonator. One of the main benefits of using ring resonators in biosensing is the small volume of sample specimen required to obtain a given spectroscopy [40, 41]. These devices have also proved useful as single photon sources for quantum information experiments [42, 43].

MEMS ring resonators are distributed-parameter systems; that means they have an infinite number of eigenfrequencies known as modes of the resonator. Each mode can be represented by an equivalent lumped model, but this approach has limited accuracy. Using a mass-spring-damper model can limit the excitation frequency to be close enough to the natural frequency of the system. The choice of modeling a given system as discrete or continuous depends on the purpose of the analysis and the expected accuracy of the results. Since accuracy is a key parameter in the analysis of MEMS resonators, a continuous model is preferable.

Motivated by the need of understanding the dynamical behavior of microring resonators, the dynamic analysis of a vibrating MEMS ring resonator is presented in this work. This paper is organized as follows. The problem formulation and the mathematical model are introduced in section 2. Based on the minimization of the Hamiltonian, the governing nonlinear equation of motion for the in-plane vibration of the ring considering the effect of van der Waals and Casimir forces is analytically derived. In section 3, the Galerkin method and the discretized equation of motion are introduced. Then the shooting method is applied to capture the periodic solutions corresponding to each excitation frequency. Section 4 reports the frequency response curves of the microring near the fundamental resonance frequency and discusses its behavior for each ratios of inner radius-to-outer radius. Finally, section 5 summarizes the results and provides concluding remarks.

## 2. MODELING

Our model presents the perspective-view schematic of a spoke-supported ring actuated by a DC load $V_{DC}$, and a harmonic load $V_{AC} Cos(\Omega t)$. As shown, this device consists of a ring structure constructed in doped polydiamond, with thickness $h$, inner radius $R_{in}$, and outer radius $R_{out}$, suspended by spokes emanating from an anchored stem self-aligned to the very center of the structure.

Considering the displacement field in the cylindrical coordinate system as $\mathbf{u}=u_r e_r + u_\theta e_\theta + u_z e_z$, the strain–displacement relations are given by [44, 45]:

$$\varepsilon_r = \frac{\partial u_r}{\partial r} \tag{1}$$

$$\varepsilon_\theta = \frac{1}{r}\frac{\partial u_\theta}{\partial \theta} + \frac{u_r}{r} \tag{2}$$

$$\gamma_{r\theta} = \frac{\partial u_\theta}{\partial r} + \frac{1}{r}\frac{\partial u_r}{\partial \theta} - \frac{u_\theta}{r} \tag{3}$$

where $\varepsilon$ and $\gamma$ are normal and shear strains respectively. Substituting for the strains in Hooke's law gives [44-46]:

$$\sigma_r = (\varepsilon_r + \nu\varepsilon_\theta)\frac{E}{1-\nu^2} = \left(\frac{\partial u_r}{\partial r} + \frac{\nu}{r}\frac{\partial u_\theta}{\partial \theta} + \frac{\nu u_r}{r}\right)\frac{E}{1-\nu^2} \tag{4}$$

$$\sigma_\theta = (\varepsilon_\theta + \nu\varepsilon_r)\frac{E}{1-\nu^2} = \left(\frac{1}{r}\frac{\partial u_\theta}{\partial \theta} + \frac{u_r}{r} + \nu\frac{\partial u_r}{\partial r}\right)\frac{E}{1-\nu^2} \tag{5}$$

$$\tau_{r\theta} = G\gamma_{r\theta} = G\left(\frac{\partial u_\theta}{\partial r} + \frac{1}{r}\frac{\partial u_r}{\partial \theta} - \frac{u_\theta}{r}\right) \tag{6}$$

where $\sigma$, $\tau$, $\nu$, $E$ and $G$ are normal stress, shear stress, Poisson's ratio, Young's modulus and shear modulus, respectively. Hamilton's principle is applied to derive the equation of motion [47]. Considering the in-plane motion, the kinetic energy is given by [45]:

$$K = \frac{1}{2}\int_0^{2\pi}\int_{R_{in}}^{R_{out}} \rho\left[\left(\frac{\partial u_r}{\partial t}\right)^2 + \left(\frac{\partial u_\theta}{\partial t}\right)^2\right] h r dr d\theta \tag{7}$$

and the potential energy is given by [45]:

$$U = \frac{1}{2}\int_0^{2\pi}\int_{R_{in}}^{R_{out}} (\sigma_r \varepsilon_r + \sigma_\theta \varepsilon_\theta + \tau_{r\theta}\gamma_{r\theta}) h r dr d\theta$$

$$= \frac{1}{2}\int_0^{2\pi}\int_{R_{in}}^{R_{out}} \left\{ \begin{array}{l} \left(\frac{\partial u_r}{\partial r} + \frac{v}{r}\frac{\partial u_\theta}{\partial \theta} + v\frac{u_r}{r}\right)\frac{E}{1-v^2}\frac{\partial u_r}{\partial r} \\ + \left(\frac{1}{r}\frac{\partial u_\theta}{\partial \theta} + \frac{u_r}{r} + v\frac{\partial u_r}{\partial r}\right)\frac{E}{1-v^2}\left(\frac{1}{r}\frac{\partial u_\theta}{\partial \theta} + \frac{u_r}{r}\right) \\ + G\left(\frac{\partial u_\theta}{\partial r} + \frac{1}{r}\frac{\partial u_r}{\partial \theta} - \frac{u_\theta}{r}\right)^2 \end{array} \right\} h r dr d\theta \tag{8}$$

Due to symmetry of both excitation and boundary conditions, in the radial-contour mode the displacement field does not depend on $\theta$; accordingly ($u_r = u_r(r,t)$) and the kinetic and potential energies reduce to:

$$K = \frac{1}{2}\int_0^{2\pi}\int_{R_{in}}^{R_{out}} \rho\left(\frac{\partial u_r}{\partial t}\right)^2 h r dr d\theta \tag{9}$$

and,

$$U = \frac{E}{2(1-v^2)}\int_0^{2\pi}\int_{R_{in}}^{R_{out}} \left\{\left(\frac{\partial u_r}{\partial r}\right)^2 + 2v\frac{u_r}{r}\frac{\partial u_r}{\partial r} + \frac{u_r^2}{r^2}\right\} h r dr d\theta \tag{10}$$

The work of the electrostatic force is expressed as:

$$w_e = \int_0^{2\pi}\int_0^{u_r(R_{out},t)} \frac{\varepsilon_0 R_{out} h}{2(d_0 - \zeta)^2}(V_{DC} + V_{AC}\cos(\Omega t))^2 d\zeta d\theta \tag{11}$$

where $\varepsilon_0$ is the permittivity of vacuum and $\zeta$ is a dummy parameter. Besides the electrostatic force, van der Waals and Casimir forces also play important roles when the ratio of the gap to the length is sufficiently small [48]. In the same way, the work of van der Waals force is given as:

$$w_1 = \int_0^{2\pi}\int_0^{u_r(R_{out},t)} \frac{R_{out} h A}{6\pi(d_0 - \zeta)^3} d\zeta d\theta \tag{12}$$

where $A$ is the Hamaker constant which for the most condensed phases are found to lie in the range (0.4-4) $\times 10^{-19}$ J [49]. For the polydiamond $A = 2.2 \times 10^{-19}$ J [50, 51]. The work of Casimir force is:

$$w_2 = \int_0^{2\pi}\int_0^{u_r(R_{out},t)} \frac{R_{out} h \pi^2 \hbar c}{240(d_0 - \zeta)^4} d\zeta d\theta \tag{13}$$

where $\hbar = 1.055 \times 10^{-34}$ Js is the Planck's constant divided by $2\pi$ and $c = 2.998 \times 10^8$ m/s is the speed of light [52].

The governing partial differential equation of motion is obtained by the minimization of the Hamiltonian using the variational principle as [53]:

$$\int_{t_1}^{t_2} \left(\delta K - \delta U + \delta w_e + \delta w_n\right) dt = 0, \qquad n = 1, 2 \tag{14}$$

The index, $n$, is 1 and 2 for the van der Waals and Casimir forces, respectively. Using the Dirac delta function, the governing equation reduces to:

$$2\pi\rho h \frac{\partial^2 u_r}{\partial t^2} - \frac{2\pi h E}{1-v^2}\left(\frac{\partial^2 u_r}{\partial r^2} + \frac{1}{r}\frac{\partial u_r}{\partial r} - \frac{u_r}{r^2}\right) = \frac{\delta(r-R_{out})\pi\varepsilon_0 R_{out} h}{r(d_0 - u_r)^2}\left(V_{DC} + V_{AC}\cos(\Omega t)\right)^2 + f_n \tag{15}$$

where:

$$\begin{cases} f_n = \dfrac{\delta(r-R_{out}) A R_{out} h}{3r(d_0 - u_r)^3}, & n = 1 \\[2mm] f_n = \dfrac{\delta(r-R_{out}) R_{out} h \pi^3 \hbar c}{120 r (d_0 - u_r)^4}, & n = 2 \end{cases} \tag{16}$$

subjected to the following boundary condition:

$$\left(r\frac{\partial u_r}{\partial r} + v u_r\right)\Bigg|_{R_{in}}^{R_{out}} = 0 \tag{17}$$

To obtain the governing differential equation of motion in its non-dimensional form the following nondimensional parameters are introduced:

$$\beta = \frac{R_{in}}{R_{out}}, \quad \hat{r} = \frac{r}{R_{out}}, \quad \hat{u}_r = \frac{u_r}{d_0}, \quad \hat{t} = \frac{t}{T}, \quad \hat{\Omega} = \Omega T, \quad T = R_{out}\sqrt{\frac{\rho(1-v^2)}{E}} \tag{18}$$

Substituting Eq. (18) in Eq. (15), and considering viscous damping effect due to the squeeze film damping and dropping the hats, the non-dimensional differential equation of the motion reduces to:

$$\frac{\partial^2 u_r}{\partial t^2} + C_{damping}\frac{\partial u_r}{\partial t} - \frac{\partial^2 u_r}{\partial r^2} - \frac{1}{r}\frac{\partial u_r}{\partial r} + \frac{u_r}{r^2} = \alpha_1\frac{\delta(r-1)\left(V_{DC} + V_{AC}\cos(\Omega t)\right)^2}{r(1-u_r)^2} + F_n \tag{19}$$

where the following non-dimensional boundary condition holds:

$$\left( r \frac{\partial u_r}{\partial r} + v u_r \right)\Bigg|_{r=\beta}^{r=1} = 0 \tag{20}$$

where $C_{\text{damping}}$ is the nondimensional damping coefficient and,

$$\begin{cases} F_n = \alpha_2 \dfrac{\delta(r-1)}{r(1-u_r)^3}, & n = 1 \\ F_n = \alpha_3 \dfrac{\delta(r-1)}{r(1-u_r)^4}, & n = 2 \end{cases} \tag{21}$$

$$\alpha_1 = \frac{R_{out}^2 (1-v^2)\varepsilon_0}{2Ed_0^3}, \quad \alpha_2 = \frac{R_{out}^2 (1-v^2)A}{6\pi E d_0^4}, \quad \alpha_3 = \frac{R_{out}^2 (1-v^2)\pi^2 \hbar c}{240 E d_0^5}$$

We need to solve the equation of motion and its boundary condition with an appropriate method to obtain the frequency response of the system. Table 1 represents the geometrical and material properties of the case study.

Table 1: Parameters of numerical calculation.

| Class | Parameter | Symbol | Values | Units |
|---|---|---|---|---|
| **Physical Constants** | Hamaker Constant of Polydiamond | $A$ | $2.2 \times 10^{-19}$ | J |
| | Planck's Constant Divided by $2\pi$ | $\hbar$ | $1.055 \times 10^{-34}$ | Js |
| | Speed of Light | $c$ | $2.988 \times 10^8$ | m/s |
| | Permittivity of Vacuum | $\varepsilon_0$ | $8.85 \times 10^{-12}$ | $C^2.N^{-1}.m^{-2}$ |
| **The Polydiamond Ring** | Ring Thickness | $h$ | 2.1 | μm |
| | Young's Modulus | $E$ | 1198 | GPa |
| | Poisson's Ratio | $v$ | 0.0691 | - |
| | Density | $\rho$ | 3500 | Kg/m$^3$ |
| | Acoustic Velocity | $v_p$ | 18500 | m/s |
| **The Gap** | Electrode-to-Ring Gap | $d_0$ | 40.0 | nm |

## 3. NUMERICAL SOLUTION

The microring displacement is approximated as:

$$u_r(r,t) = \sum_{i=1}^{N} q_i(t)\varphi_i(r) \tag{22}$$

where $\varphi_i(r)$ and $q_i(t)$ are the linear mode shapes and the generalized coordinates, respectively, and $N$ is the number of assumed modes.

### 3.1. Mode Shape and Fundamental Natural Frequency

In this section the fundamental procedure for the determination of natural frequency of a linearly-behaving microring and its corresponding mode shape through solving the so-called eigenvalue problem is presented. Toward this we refer back to Eq. (19) and set the forcing and damping terms equal to zero. Substituting Eq. (22), considering the first mode and simplifying the terms, Eq. (19) reduces to:

$$r^2 \frac{\partial^2 \varphi(r)}{\partial r^2} + r \frac{\partial \varphi(r)}{\partial r} + \left(\lambda^2 r^2 - 1\right)\varphi(r) = 0 \tag{23}$$

which is in the form of Bessel's equation of order 1 with the parameter $\lambda$ (the first non-dimensional natural frequency of the resonator). The solution of Eq. (23) is given by:

$$\varphi(r) = M J_1(\lambda r) + N Y_1(\lambda r) \tag{24}$$

where $J_1$ and $Y_1$ are the Bessel functions of the first and the second kind of order 1, and $M$ and $N$ are arbitrary constants. Using Eq. (24), Eq. (20) is expressed as:

$$\begin{aligned}[\beta \lambda Y_0(\lambda \beta) - (1-\nu) Y_1(\lambda \beta)][\lambda J_0(\lambda) - (1-\nu) J_1(\lambda)] \\ -[\beta \lambda J_0(\lambda \beta) - (1-\nu) J_1(\lambda \beta)][\lambda Y_0(\lambda) - (1-\nu) Y_1(\lambda)] = 0 \end{aligned} \tag{25}$$

Eq. (25) is the frequency equation for the in-plane vibration of a ring. Solving this equation for $\lambda$, fundamental frequency and the appertaining mode shape of the ring are obtained as:

$$f_0 = \frac{\lambda}{2\pi R_{out}} \sqrt{\frac{E}{\rho(1-\nu^2)}} \quad \text{(Hz)} \tag{26}$$

and,

$$\varphi(r) = J_1(\lambda r) - \left[\frac{\lambda J_0(\lambda) - (1-\nu) J_1(\lambda)}{\lambda Y_0(\lambda) - (1-\nu) Y_1(\lambda)}\right] Y_1(\lambda r) \tag{27}$$

### 3.2. Solving the equation of motion

Substituting Eq. (22) into Eq. (19), and applying the Galerkin discretization method based on the orthogonality of Bessel functions, the reduced order differential equation of the microring reduces to:

$$\sum_{i=1}^{M}\ddot{q}_i(t)M_{ij} + \sum_{i=1}^{M}\dot{q}_i(t)C_{ij} + \sum_{i=1}^{M}q_i(t)K_{ij} = \int_{\beta}^{1}\frac{\alpha_1\delta(r-1)(V_{DC}+V_{AC}\text{Cos}(\Omega t))^2\varphi_j(r)}{\left(1-\sum_{i=1}^{M}q_i(t)\varphi_i(r)\right)^2}dr + \Pi_n \qquad (28)$$

where

$$\begin{cases} \Pi_n = \int_{\beta}^{1}\dfrac{\alpha_2\delta(r-1)\varphi_j(r)}{\left(1-\sum_{i=1}^{M}q_i(t)\varphi_i(r)\right)^3}dr, & n=1 \\[2mm] \Pi_n = \int_{\beta}^{1}\dfrac{\alpha_3\delta(r-1)\varphi_j(r)}{\left(1-\sum_{i=1}^{M}q_i(t)\varphi_i(r)\right)^4}dr, & n=2 \end{cases} \qquad (29)$$

$$M_{ij} = \int_{\beta}^{1}\varphi_i(r)\varphi_j(r)rdr$$

$$C_{ij} = C_{\text{damping}}\int_{\beta}^{1}\varphi_i(r)\varphi_j(r)rdr$$

$$K_{ij} = -\int_{\beta}^{1}\varphi_i''(r)\varphi_j(r)rdr - \int_{\beta}^{1}\varphi_i'(r)\varphi_j(r)dr + \int_{\beta}^{1}\frac{1}{r}\varphi_i(r)\varphi_j(r)dr$$

The prime denotes differentiation with respect to *r* and the overdot denotes differentiation with respect to *t*. We have applied the shooting method to capture the periodic solutions of Eq. (28). The shooting technique as a powerful method to handle boundary value problems, can be used to capture the periodic orbits and accordingly the steady state responses of both autonomous and nonautonomous systems [54].

## 4. RESULTS AND DISCUSSIONS

In this section, we will study and analyze the static and dynamic behavior of a micromechanical radial-contour mode ring resonator. To find the static response, we set all the time dependent terms in Eq. (19) equal to zero. The influence of van der Waals (vdW) and Casimir forces on the frequency response of the model is also studied. Figure 1 compares between the radial displacement at the edge of the ring resonator with and without considering the van der Waals and Casimir effects. These forces are intending to increase the displacement of the edge. As the gap becomes smaller the effect of vdW and Casimir also increases.

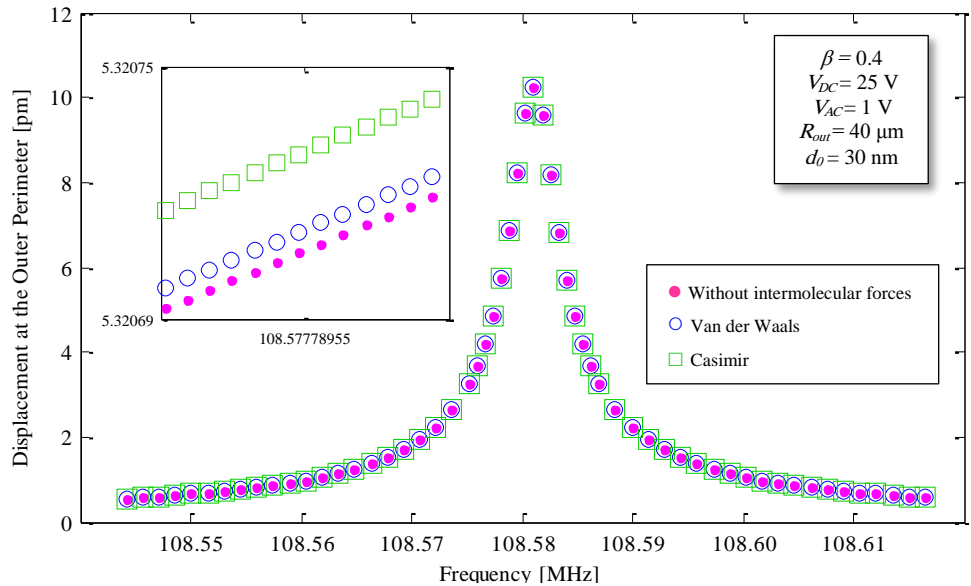

Figure 1: Effect of vdW and Casimir forces on the frequency response curve.

The higher the quality factor, the more the resonator efficiency. This fact is illustrated in Figure 2, where the $Q$ changes from 11,573 to 57,867. In this figure the variation of the radial displacement with the quality factor is presented. It can be concluded that for the ring resonators the amount of the quality factor does not qualitatively affect the frequency response curve but changes the amplitude of the periodic solutions.

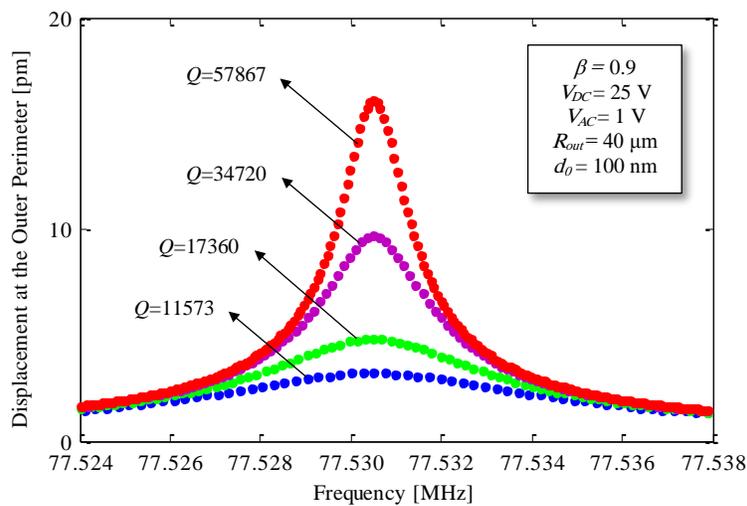

Figure 2: Frequency response curves showing the effect of quality factor ($Q$).

The main advantage of these in-plane vibrating ring resonators is their weak nonlinearity, which can be seen in the frequency responses of this section. For out-of-plane capacitors, nonlinearity limits the controlled travel range of actuators and can result in unexpected collapse, short circuit, stiction, and functional failure of sensors [54]. The theoretical results presented in this paper show very good agreement with the experimental results [35, 39, 55] and simulations [56].

## 5. CONCLUSION

We presented a method to model and obtain the dynamic response of an electrostatically actuated microring resonator. The governing equation is analytically derived incorporating the effects of electrostatic, van der Waals and Casimir forces. Using a Galerkin method, the discretized equation of motion is introduced and numerically integrated over time. The influences of intermolecular forces such as van der Waals and Casimir on the dynamic behavior of the resonator are investigated. The effects of increased forcing and decreased damping on the frequency response are investigated. This work provides new theoretical insights on ring resonators and analytical solutions that can be used for the design and optimization of ring resonators under different conditions.